\documentclass[letterpaper,11pt]{article}

\usepackage[top=1.0in,bottom=1.0in,left=1.0in,right=1.0in]{geometry}

\usepackage{url}

\begin{document}

\title{Comment on ``Optimal prior for Bayesian inference in a 
constrained parameter space'' by S. Hannestad and T. Tram,
 arXiv:1710.08899}
\author{Robert D. Cousins\thanks{cousins@physics.ucla.edu}\\
Dept.\ of Physics and Astronomy, 
University of California, Los Angeles,
California 90095 USA}
 
\date{February 20, 2019}

\maketitle

\noindent {\bf Abstract:} The Jeffreys prior for a constrained part of
a parameter space is the same as that for the unconstrained space,
contrary to the assertions of Hannestad and Tram.

\bigskip
\noindent
An arXiv post by Hannestad and Tram~\cite{hannestad2017} states, 
\begin{quote}
Under the assumption that the likelihood is a Gaussian distribution,
the Jeffreys prior is a constant, i.e. flat. However, if one parameter
is constrained by physical considerations, the Gaussian approximation
fails and the flat prior is no longer the Jeffreys prior\dots In this
paper we compute the correct Jeffreys prior for a multivariate normal
distribution constrained in one dimension\dots
\end{quote}
When this was pointed out to me last week, I thought this odd since I
have long been teaching (including recently in Section 5.3 of
Ref.~\cite{cousins2018}), that such a constraint does not modify the
Jeffreys prior, based on my memory of Jim Berger's
book~\cite{berger1985}.  Since this specific result of
Ref.~\cite{hannestad2017} is being cited without
criticism~\cite{Casas:2017wjh,Gariazzo:2018pei,deSalas:2018idd,Reischke:2018ooh,Diacoumis:2018ezi},
it is worth opening Berger's book and finding on page 89:
\begin{quote}
One important feature of the Jeffreys noninformative prior is that it
is not affected by a restriction on the parameter space.  Thus, if it
is known in Example 5 that $\theta>0$, the Jeffreys noninformative
prior is still $\pi(\theta)=1$ (on $(0,\infty)$, of course).  This is
important, because one of the situations in which noninformative
priors prove to be extremely useful is when dealing with restricted
parameter spaces (see Chapter 4). In such situations we will,
therefore, simply assume that the noninformative prior is that which
is inherited from the unrestricted parameter space.
\end{quote}

\noindent
(Example 5 is for a general location parameter, which includes the
Gaussian case of Ref.~\cite{hannestad2017}.)  As the Jeffreys prior is
problematic with more than one parameter, ``reference priors'' were
developed by Bernardo and Berger.  According to
Ref.~\cite{bergersun1998}, in general they share this important
feature:
\begin{quote}
Another common type of partial information is constraints on the
parameter space. This is typically easily handled, however, in that
reference priors for a constrained space are almost always just the
unconstrained reference prior times the indicator function on the
constrained space.
\end{quote}

\noindent
It appears that Hannestad and Tram have fallen into a trap that I
discuss in Section 6.9 of Ref.~\cite{cousins2018}, namely the ``Famous
confusion re Gaussian $p(x| \mu)$ where $\mu$ is mass $\ge0$.''  That
is, they view the sampled value $x$ (called $q$ in their paper) as an
estimate of $\mu$ and thus consider only $x\ge0$.  In truth, there is
nothing anomalous about a sampled $x<0$, and the computation of the
expectation value over $x$ should not be restricted to $x\ge0$.  Once
this is understood (that restrictions in the parameter space do not
restrict the sample space just because one may like to think of $x$ as
the estimate or ``measured value'' of the parameter), the quoted
passage from Berger's book becomes obvious.  If one still has trouble
distinguishing sampled value of $x$ from point estimate of $\mu$,
Section 6.4 of Ref.~\cite{cousins2018} has an instructive example.  As
noted at the end of Section 6.8, the confusion is also avoided if $x$
and $\mu$ have different units or dimensions.

In any case, it would seem that this result of Hannestad and Tram is
contradicted by well-established professional statistics literature.

\section*{Acknowledgments}
I think Maurizio Pierini for pointing me to Ref.~\cite{hannestad2017}.
This work was partially supported by the U.S.\ Department of Energy
under Award Number {DE}--{SC}0009937.

\end{document}